\documentclass[sigplan,10pt,screen, letterpaper]{acmart}

\usepackage[utf8]{inputenc}
\usepackage{graphicx}
\usepackage{textcomp}
\usepackage{tikz}
\usepackage{float}
\usepackage{subfigure}
\usepackage{amsmath}
\usepackage{color}
\usepackage{framed}
\usepackage{import}
\usepackage{balance}

\usepackage{lipsum}                     
\usepackage{xargs} 
\usepackage[listings,skins]{tcolorbox}

\usepackage{url}
\usepackage{hyperref}

\definecolor{lightgray}{rgb}{.9,.9,.9}
\definecolor{darkgray}{rgb}{.4,.4,.4}
\definecolor{purple}{rgb}{0.65, 0.12, 0.82}

\lstdefinelanguage{JavaScript}{
  keywords={typeof, new, true, false, catch, function, return, null, catch, switch, var, if, in, while, do, else, case, break},
  keywordstyle=\color{blue}\bfseries,
  ndkeywords={class, export, boolean, throw, implements, import, this},
  ndkeywordstyle=\color{darkgray}\bfseries,
  identifierstyle=\color{black},
  sensitive=false,
  comment=[l]{//},
  morecomment=[s]{/*}{*/},
  commentstyle=\color{purple}\ttfamily,
  stringstyle=\color{red}\ttfamily,
  morestring=[b]',
  morestring=[b]"
}

\lstdefinelanguage{IR}{
  keywords={StackCheck, LdaSmi, Start, LdaNamedProperty, Add, Return, AddSmi},
  keywordstyle=\color{blue}\bfseries,
  ndkeywords={class, export, boolean, throw, implements, import, this},
  ndkeywordstyle=\color{darkgray}\bfseries,
  identifierstyle=\color{black},
  sensitive=false,
  comment=[l]{//},
  morecomment=[s]{/*}{*/},
  commentstyle=\color{purple}\ttfamily,
  stringstyle=\color{red}\ttfamily,
  morestring=[b]',
  morestring=[b]"
}

\lstdefinelanguage{IR_JS}{
  keywords={StackCheck, LdaSmi, Start, LdaNamedProperty, Add, Return, AddSmi, typeof, new, true, false, catch, function, return, null, catch, switch, var, if, in, while, do, else, case, break},
  keywordstyle=\color{blue}\bfseries,
  ndkeywords={class, export, boolean, throw, implements, import, this},
  ndkeywordstyle=\color{darkgray}\bfseries,
  identifierstyle=\color{black},
  sensitive=false,
  comment=[l]{...},
  morecomment=[s]{/*}{*/},
  commentstyle=\color{purple}\ttfamily,
  stringstyle=\color{red}\ttfamily,
  morestring=[b]',
  morestring=[b]"
}

\lstdefinelanguage{EV}{
  keywords={,},
  keywordstyle=\color{blue}\bfseries,
  ndkeywordstyle=\color{darkgray}\bfseries,
  sensitive=false
}

\startPage{1}

\begin{document}

\title[Scalable Comparison of JavaScript V8 Bytecode Traces]{Scalable Comparison of JavaScript V8 Bytecode Traces}         


\author{Javier Cabrera Arteaga}
\orcid{0000-0001-9399-8647}             
\affiliation{
  \institution{KTH Royal Institute of Technology}            
  \city{Stockholm}
  \postcode{11428}
  \country{Sweden}                    
}
\email{javierca@kth.se}          

\author{Martin Monperrus}
\orcid{0000-0003-3505-3383}             
\affiliation{
  \institution{KTH Royal Institute of Technology}           
  \city{Stockholm}
  \postcode{11428}
  \country{Sweden}                   
}
\email{martin.monperrus@csc.kth.se}         

\author{Benoit Baudry}
\orcid{0000-0002-4015-4640}             

\affiliation{
  \institution{KTH Royal Institute of Technology}           
  \city{Stockholm}
  \postcode{11428}
  \country{Sweden}                    
}
\email{baudry@kth.se}         

\begin{abstract}

The comparison and alignment of runtime traces are essential, e.g., for semantic analysis or debugging. However, naive sequence alignment algorithms cannot address the needs of the modern web: (i) the bytecode generation process of V8 is not deterministic; (ii) bytecode traces are large. 

We present STRAC, a scalable and extensible tool tailored to compare bytecode traces generated by the V8 JavaScript engine. Given two V8 bytecode traces and a distance function between trace events, STRAC computes and provides the best alignment. The key insight is to split access between memory and disk. STRAC can identify semantically equivalent web pages and is capable of processing huge V8 bytecode traces whose order of magnitude matches today's web like \url{https://2019.splashcon.org}, which generates approx. 150k of V8 bytecode instructions.

\end {abstract}


\maketitle

\newcommand{\ngram}{\textbf{n}-gram }
\newcommand{\ngrams}{\textbf{N}-grams }
\newcommand{\xgram}[1]{\textbf{{#1}}-gram }

\newcommand{\red}[1]{{\color{red} #1 }}
\newcommand{\blue}[1]{{\color{blue} #1 }}

\newcommand{\hiddensite}{{\url{https://kth.se}}}

\newcommand{\repo}{ \sloppy \url{https://github.com/KTH/STRAC}
 }

\lstdefinestyle{mycodeStyle}{
  numbers=left,
  stepnumber=1,
  numbersep=10pt,
  tabsize=4,
  showspaces=false,
  showstringspaces=false
}

\vspace{0.7cm}
\section{Introduction}
Runtime traces record the execution of programs.
This information captures the dynamics of programs and can be used to determine semantic similarity \cite{6280299}, to detect abnormal program behavior \cite{jiang_multiresolution_2007-1}, to check refactoring correctness \cite{Ramos} or to infer execution models \cite{ beschastnikh_leveraging_2011}. 
In many cases, this is achieved by comparing execution traces, e.g. comparing the traces of the original program and the refactored one.
The comparison of program traces can be based on information retrieval \cite{6976081}, tree differencing \cite{8327321, DBLP:journals/corr/abs-1708-03786} and sequence alignment \cite{Kim2017, Churchill}. In this paper, we focus on the latter, in order to compare sequences of V8 bytecode instructions resulting from the execution of JavaScript code.


V8 is an open source, high-performance JavaScript engine. For debugging purposes, it provides powerful facilities to export page execution information \cite{noauthor_run_nodate}, including intermediate internal bytecode called the V8 bytecode \cite{ignition}.

Due to the dynamic nature of the Web, we observe that the bytecode generation process of V8 is not deterministic. For example, visiting the same page several times results in different V8 bytecode traces every time. This non-determinism is a key challenge for sequence alignment approaches, even if they perform well on deterministic program traces \cite{Kargen:2017:TRI:3155562.3155608}. 
Besides, V8 bytecode traces are large. Naive sequence alignment algorithms are time and space quadratic on trace sizes and do not scale to V8 bytecode traces. To illustrate this scaling problem, let us consider a simple query to \url{https://2019.splashcon.org}: it generates between 139555 and 162558 V8 bytecode instructions, and aligning two traces of such size, requires approximately 150GB of memory\footnote{In this paper, memory means RAM.}. This memory requirement is not realistic for trace analysis tasks on developer's personal computers or servers. The key challenge that we address in this work is to provide a trace comparison tool that scales to V8 bytecode traces.

In this paper, we present STRAC (Scalable Trace Comparison), a scalable and extensible tool tailored to compare bytecode traces from the V8 JavaScript engine. STRAC implements an optimized version of the DTW algorithm \cite{NEEDLEMAN1970443}. Given two V8 bytecode traces and a distance function between trace events, STRAC computes and provides the best alignment. The key insight is to split access between memory and disk.

Our experiments compare STRAC with 6 other publicly-available implementations of DTW. The comparison involves $100$ pairs of V8 bytecode traces collected over 6 websites.
Our experimental results show that 1) STRAC can identify semantically equivalent web pages and 2) STRAC is capable of processing big V8 bytecode traces whose order of magnitude matches today's web. 

To sum up, our contributions are:

\begin{itemize}
    \item An analysis of the challenges for analyzing browser traces, due to the JavaScript engine internals and the randomness of the environment. We explain and show examples of how the same browser query can generate two different V8 bytecode traces. 

    \item A tool called STRAC that implements the popular alignment algorithm DTW in a scalable way, publicly available at \repo.  

    \item A set of experiments comparing 100 V8 bytecode traces collected over 6 real world websites:\url{google.com}, \url{kth.se}, \url{github.com}, \url{wikipedia.org}, \url{2019.splashcon.org} and  \url{youtube.com}. Our experiments show that STRAC copes with the non-deterministic traces and is significantly faster than state-of-the-art tools.
\end{itemize}

The paper is structured as follows. First we introduce a background of V8 bytecode generation non-determinism and the formalisms used in our work (\autoref{back}). Then follows with technical insights to implement STRAC (\autoref{strac}), research question formulation, experimental results with a discussion about them (\autoref{exp}). We then present related work (\autoref{rw}) and conclude (\autoref{conclusion}).

\section{Background}
\label{back}

In this section we discuss the key insights behind the non-determinism of the V8 bytecode generation process, as well as the foundations of the DTW alignment algorithm.

\subsection{Browser Traces}

Our dynamic analysis technique is evaluated with V8 bytecode \cite{noauthor_v8_nodate-1}.
In this subsection, we describe how the V8 engine generates bytecode trace. We collect such traces to evaluate our trace comparison tool. In this work, we use the term "V8 bytecode trace" to refer to the result of executing V8 with the \textit{--print-bytecode} flag \cite{noauthor_run_nodate}. 

\subsubsection{V8 Bytecode Generation}
\label{v8}


The V8 engine compiles JavaScript source code to an intermediate representation called ``V8 bytecode''. This is done to increase execution performance. The V8 engine parses and compiles every JavaScript code declaration present in HTML pages into a bytecode representation, composed by function declarations, like the one shown in \autoref{bytecode_figure}. These function declarations came from V8 builtin JavaScript code and external JavaScripts.

V8's bytecode interpreter is a register machine \cite{ignition16}. \autoref{bytecode_figure} shows a JavaScript code and its bytecode translation. Each bytecode operator specifies its inputs and outputs as register operands. V8 has 180 different bytecode operators. 


The bytecode translation is lazy, i.e. V8 tries to avoid generating code it "thinks" might not be executed.  Consequently, a function that is not called will not be compiled \cite{noauthor_blazingly_nodate}. For example, removing line 2 in the top listing of \autoref{bytecode_figure} would prevent the compilation of bytecode for the function declared in line 1. This behavior has an impact on the collected traces.

\begin{figure}[H]

  \lstset{language=JavaScript,
                  style=mycodeStyle,
                  basicstyle=\small\ttfamily,
                  columns=fullflexible,
                  breaklines=true,
                  linewidth=0.97\linewidth,
                  postbreak=\mbox{\textcolor{red}{$\hookrightarrow$}\space}}

      \begin{lstlisting}[]
  function plusOne(a){ return a.value + 1; }
  plusOne( {value : 2018} );  
  \end{lstlisting}
  \rule{0.94\linewidth}{0.1mm}
  \lstset{language=IR,
                  style=mycodeStyle,
                  basicstyle=\small\ttfamily,
                  columns=fullflexible,
                  breaklines=true,
                  postbreak=\mbox{\textcolor{red}{$\hookrightarrow$}\space}}
  \begin{lstlisting}[]
  [generated bytecode for function: plusOne]
  Parameter count 2
  Register count 0
  Frame size 0
  30 E> 0x1373c709b6 @    0 : a5 00 00 00       StackCheck 
  56 S> 0x1373c709b7 @    1 : 28 02 00 01       LdaNamedProperty a0, [0], [1]
  62 E> 0x1373c709bb @    5 : 40 01 00 00       AddSmi [1], [0]
  66 S> 0x1373c709be @    8 : a9 00 00 00       Return 
  \end{lstlisting}

\caption{Example of a JavaScript function and its corresponding V8 bytecode instructions.}
\label{bytecode_figure}
\end{figure}

We have observed that V8 bytecode is resilient to script minification and static code-obfuscation techniques. Therefore, we believe that aligning such low-level representations could prove to be a useful aid in many program analysis tasks, such as code similarity study and malware analysis.

\subsubsection{Non-Determinism in Browser Traces}
\label{non-determinism}

\begin{figure}[htbp]
  \centerline{
  \import{diagrams/}{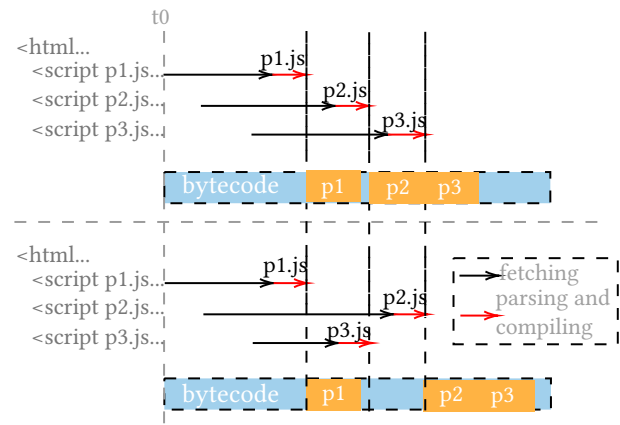}}
  \caption{Illustration of two different script fetching and compiling traces for the same browser query.}
  \label{parsing}
  \end{figure}

Interestingly, browsers are fundamentally non deterministic, depending on web server availability, current workload, and DNS caches through the network.
Let us look at the example illustrated in \autoref{parsing}. 
It shows what happens when fetching a web page, which contains 3 scripts.
The top and bottom parts illustrate, for the same page, two different executions. Dashed border rectangles represent complete bytecode generation traces. The blue spaces in the bar are V8 common builtin bytecode, which is systematically generated in all browser requests. Orange rectangles illustrate declared page scripts compilations. The complete bytecode trace is the union of both generated bytecodes, builtin V8 and page declared scripts. In the first case at the top of \autoref{parsing}, the scripts are fetched and compiled in the same order they are declared. In the second case, at the bottom, \textbf{p3.js} is carried and compiled first, before \textbf{p2.js} due to a possible network delay. 
 However, V8's compiler will put all scripts compilations in the same order they are declared in the HTML page. The final result is two semantically equivalent bytecode compilations, where script blocks may not be strictly placed in the same position.
 
 

The slight differences that occur in the final bytecode for same browser queries motivate us to provide an efficient tool for traces alignment: traces where events occur in different orders but that have the same semantics must be considered as equivalent. The order of events should not confuse the trace comparison tool.




%


\subsubsection{DTW Algorithm}
\label{dtw_algorithm}

The DTW algorithm has been introduced by Needleman and Wunsch for protein global alignment \cite{NEEDLEMAN1970443}. Global alignment means trace heads and tails are constrained to match each other in position. 
DTW is a popular technique for comparing traces in different domains, incl. software traces \cite{ Maia08usinga}.
DTW finds the best global alignment between two traces, based on a generic similarity function between trace events and gaps.

\textbf{Definition (Trace)} A trace $X$ is defined as a sequence of events.  \textbf{$X=x_1, x_2, ... x_N$} represents a trace of size $N$ where each $x_i$ is the event happening at the ith position. 

\textbf{Definition (Cost Matrix)} $D$ is a cost matrix for two traces $X$ and $Y$ of size $n$ and $m$. $D_{ij}$ stores the optimal cost alignment value for $X$ and $Y$ considered from the start up to the $ith$ and $jth$ positions respectively, that is the minimal cost of aligning $x_i$ and $y_j$ events at the same position in the final alignment.

The cost matrix is defined according to a distance function $d$ and a gap cost $\gamma$ as follows:

\vspace{0.2cm}

{\centering
$D_{0i} = \gamma*i$

\vspace{0.2cm}

$D_{j0} = \gamma*j$

$$D_{ij}= min
    \begin{cases}
    D_{i - 1j} + \gamma,\\
    D_{i j - 1} + \gamma \\
    D_{i - 1j - 1} + d(x_i, y_j)
    \end{cases}
$$
}


In every cell, the value $D_{ij}$ is the minimum cost between putting a gap in one trace and the result of evaluating the distance function between events $x_i$ and $y_j$. 

\textbf{Definition (Alignment Cost)} Given two traces $X$ and $Y$ with sizes $N$ and $M$ respectively, the alignment cost is the value stored in $D_{NM}$. 

\textbf{Definition (Alignment Difficulty)} Given two traces $X$ and $Y$ with sizes $N$ and $M$ respectively, the alignment difficulty is simply the multiplication of both sizes $N\times M$. 

\textbf{Definition (Warp Path)} The warp path is the path to go from $D_{NM}$ to the first element $D_{00}$  minimizing the cumulative cost. In general more than one path may exist. Size of warp path is $O(N + M)$.

\textbf{Definition (Aligned Trace)} An aligned trace is a trace where the warp path is applied, i.e. some gaps have been put between some events in one of both traces. 

\begin{figure}[htbp]
\import{diagrams/}{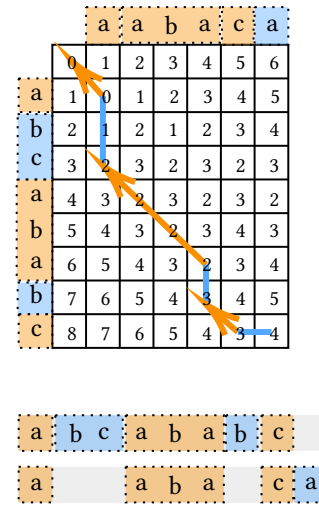}
\begin{tikzpicture}

\end{tikzpicture}

\caption{Cost matrix, warp path and applied alignment for \textbf{abcababc} and \textbf{aabaca} example traces.}
\label{cost_matrix}
\end{figure}

In \autoref{cost_matrix} we illustrate the alignment between traces \textbf{abcababc} and \textbf{aabaca} with $\gamma=1$, $d(x_i, y_j) = 2$ if $x_i \neq y_i$ and $d(x_i, y_j) = 0$ if $x_i = y_i$. The warp path is represented as the blue and orange lines going across the matrix from the top left corner to the bottom right corner. In this example, alignment cost is 4, as we can see in bottom right corner cell in \autoref{cost_matrix}.


\section{STRAC: Trace Comparison Tool for V8}
\label{strac}

STRAC is an approach to compare large traces, tailored to bytecode traces of the V8 JavaScript engine. STRAC takes as input a trace of JavaScript V8 bytecode traces collected in the browser. It produces as output, a trace alignment, and a distance measure between the two traces.
STRAC implements the DTW algorithm  presented in  \autoref{dtw_algorithm}. It is an open-source project publicly-available on \repo.
 In this section, we explain the key components and insights of STRAC to achieve scalable trace comparison.

\subsection{Challenges Addressed by STRAC}
\label{challenge}

\textbf{Non-Determinism} As shown in \autoref{non-determinism}, V8 can provide two different bytecode traces for the same web page. In this case, both traces are semantically equivalent, but the global position of code modules can vary. These variations occur as a consequence of resource management, interpreter optimizations and JavaScript code fetching from the network. 
It is challenging because it can provide
1) false positives: two traces may be considered different even when they come from the same pages;
2) false negatives: two traces may be considered the same even when they come from two different pages.

\textbf{Size} Browser traces are huge and naive trace comparison fails on such traces because of memory requirements.
For instance, aligning two traces of size 63137 and 58265 events requires a DTW cost matrix, represented as a bidimensional integer matrix, of 14.72 GB of memory.
The challenge is to make trace comparison at the scale of browser traces, with tractable memory requirements.

\subsection{DTW Distance Functions}
\label{parametrization}

The DTW algorithm has two main parameters: a distance function and a gap cost as explained in \autoref{dtw_algorithm}. The distance function between events affects the global alignment result, as we show in \autoref{similarity}. It defines the matching of two different trace instructions if these instructions have a certain level of similarity. For example, when comparing \textit{'AddSmi [0], [1]'} and \textit{'AddSmi [1], [0]'} instructions, they can be considered as similar because the \textit{AddSmi} operator is in both. 

In STRAC, we define two distance functions for bytecode instructions.

 $$ d_{Sen}(x_i, y_j) = 
      \begin{cases}
      s \text{  if $x_i$ and $y_j$ events are exactly} \\ \text{ the same bytecode instruction} \\
      c \text{  otherwise}
      \end{cases}
$$
$$ d_{Inst}(x_i, y_j) = 
      \begin{cases}
      s \text{  if $x_i$ and $y_j$ bytecode instructions} \\ \text{ share the same bytecode operator} \\
      c \text{  otherwise}
      \end{cases}
$$

Both require the identity relationship of the bytecode instruction. For V8 bytecode, based on our results (\autoref{similarity}), it seems incoherent to accept an alignment match with two different elements instead of introducing the gap. 


We now discuss the value of $\gamma$, $s$ and $c$. The cost of introducing a gap, intuitively, must be less than the cost of matching two different events, i.e. $\gamma<s$. $c$ is the value of matching two equal events, $0$.
The default values are based on our experience, $s=5$, $\gamma=1$ and $c=0$. The three are configurable.

\subsection{Buffering the Cost Matrix}
\label{buffered_section}


The key limitation of DTW is the need for a large cost matrix to retrieve the warp path. Recall our example requiring 14.72 GB in \autoref{challenge}. This means that a naive implementation can only compare small traces due to memory explosion. 

In STRAC, we solve this problem by storing the cost matrix both in memory and disk. Only the appropriate values are kept in memory. 
Our key insight is that the current value $D_{ij}$ in the cost matrix is calculated with the previous row and column, consequently, only $O(N)$ memory space is needed to compute $D_{NM}$.
Thus, STRAC only maintains the current and previous row in memory for each DTW iteration. 
After processing a row, it is saved to disk. STRAC eventually saves the complete cost matrix to disk.

For traces with lengths 63137 and 58265, instead of 14.72 GB, STRAC requires no more than 86MB of memory for the trace alignment, which represents an improvement of 99.5\% in memory consumption. 

\subsection{Retrieving the Warp Path}

In addition to the alignment cost, it is necessary to obtain the warp path in order to create and analyze the aligned traces. Recall that the aligned traces are obtained by applying the warp path on both initial traces, as we mentioned in \autoref{dtw_algorithm}. 

To retrieve the warp path from the final cost matrix, one goes backward and starts from the trace tail positions ($D_{NM}$). Cost matrix in $D_{ij}$ depends on three neighbors $D_{i-1j}$, $D_{ij-1}$ and $D_{i - 1j - 1}$.  The backtracking process finishes when the trace start is reached, i.e. when the left top corner $D_{00}$ is reached in the matrix.
In the warp path construction process, trace indices are always decreasing by one, i.e. trace events are visited only once. Therefore, in STRAC, backtracking over the final cost matrix requires only $O(N + M)$ read operations on disk, which is scalable.


\subsection{DTW Approximations}
\label{approximations}
Due to the quadratic time and space complexity of DTW, previous work has proposed approximations to speed up the alignment process. STRAC also implements two state-of-the-art DTW  approximations. We now mention these two approximations.

\textbf{Fixed Regions} Using fixed regions is a technique only to evaluate a specified region in the cost matrix  \cite{sakoe_dynamic_1978,itakura_minimum_1975,7429637, DBLP:journals}. Consequently, the globally optimal warp path will not be found if it is not entirely in the window. This improvement speeds up DTW by a constant factor, but the execution time is still O(NM). STRAC provides support for fixed regions.

\textbf{FastDTW} \footnote{The implementation mentioned in the original paper (\url{https://cs.fit.edu/~pkc/FastDTW/}) was not available at the moment of this work.} \cite{Salvador:2007:TAD:1367985.1367993} is an approximation of DTW that has a linear time and space complexity. It combines data abstraction and constraint search in the solution space. STRAC implements FastDTW. Note that, for DTW and its approximations, the default mode is the buffering mode presented in \autoref{buffered_section}.

\subsection{Recapitulation}

To sum up, STRAC is an optimized implementation of DTW and two approximations with distance functions dedicated to V8 bytecode traces and with neat handling of the cost matrix over memory and disk in order to scale.

\section{Experimental Evaluation}
\label{exp}

We assess the scalability of STRAC for V8 bytecode trace comparison with the following research questions:
\begin{itemize}
    \item RQ1 (Scalability): To what extent does STRAC scale to traces of real-world web pages? 
    \item RQ2 (Consistency): To what extent does STRAC identify similarity in semantically-equivalent traces? 
    \item RQ3 (Distance Functions): What is the effectiveness of STRAC support of different distance functions?
    
\end{itemize}

\subsection{Study Subjects}
\label{study_subjects}

Our experiment is based on tracing the home page of the following sites; 
\url{google.com}, \url{github.com}, \url{wikipedia.org}, \url{youtube.com}, four of the most visited websites, according to Alexa. We also add two sites based on personal interest: \url{2019.splashcon.org} and \url{kth.se}, the homepage of our University. All those pages use JavaScript code. The traces were generated just opening the page without any other further action.
Since the traces are non-deterministic, we collect 100 traces for the same page. This means we collect 600 traces in total.

\begin{table}[H]

  \caption {Descriptive statistics of our benchmark. The 6 sites are sorted by popularity according to the Alexa index. Example bytecodes are available in \url{https://github.com/KTH/STRAC/tree/master/STRACAlign/src/test/resources/bytecodes}.}

  \begin{tabular}{l c c} 
  
      \hline
      Site & No. scripts & Bytecode size  \\
      \hline\hline
      google.com  & 5 & 85768    \\ 
      \hline 
      youtube.com & 15  & 166626  \\ 
      \hline
      wikipedia.org& 4 & 48260  \\ 
      \hline
      github.com & 3 & 59384  \\ 
      \hline
      kth.se & 9 & 64178 \\ 
      \hline
      2019.splashcon.org & 17 & 147196  \\ 
      \hline
  \end{tabular}

\label{sizes}
\end{table}

\begin{figure}[htbp]
  \includegraphics[width=\linewidth]{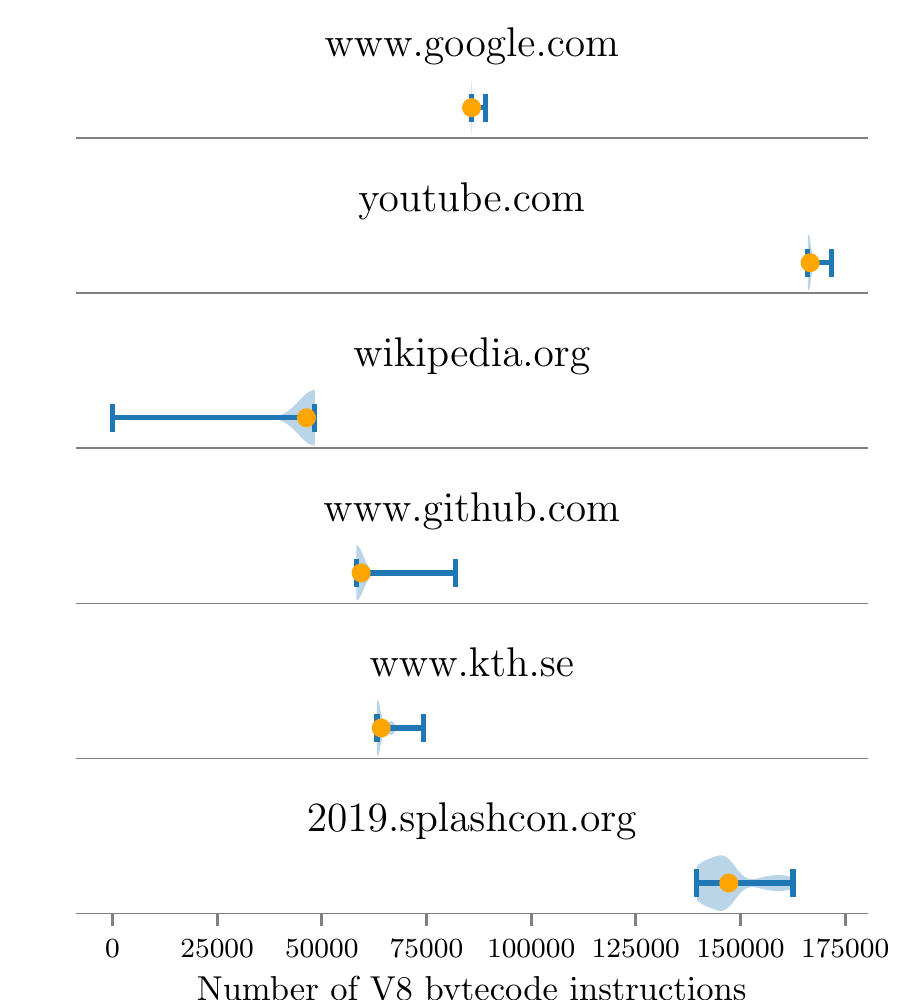}
  \caption{Variance of V8 bytecode trace size for 100 repetitions of the same query.}
  \label{v8_bytecode_distribution}
\end{figure}

\autoref{sizes}  gives an overview of the collected traces.
The first column shows the real world website names. The second and third columns indicate the number of declared scripts and the bytecode size mean value (orange dots in \autoref{v8_bytecode_distribution}) respectively.
For instance, Wikipedia loads 4 scripts and produces bytecode traces of 48260 bytecode instructions.
This value is the lowest of our benchmark.
On the contrary, for Youtube, the page declares 
15 JavaScript scripts, and V8 generates traces of 166626 bytecode instructions, and this is due to the richer features of Youtube compared to Wikipedia. In our benchmark, the bytecode traces are in the range of 48k-166k instructions.

Recall that the bytecode traces are non-deterministic even for the same page (see \autoref{non-determinism}).
We measure how many instructions are contained in each V8 bytecode trace. \autoref{v8_bytecode_distribution} illustrates the distribution of trace sizes as violin plots. 
This figure shows that there is a variance of bytecode traces for all pages (Wikipedia also has some variance but this is not shown in the figure because of the scale). 
This variance is a consequence of several stacked factors: resource management, interpreter optimization and JavaScript code fetching from the network. 
To our knowledge, this non-determinism in web traces is overlooked by research.

\subsection{Experimental Methodology}

Every trace is collected using a non-cached browser session, without plugins. This choice is motivated by two main reasons: 1) we have observed that cached scripts do not affect bytecode generation as direct network fetching does;  2) browser plugins are compiled to the same bytecode trace and in the scope of this work we are interested only in V8 bytecode traces directly generated from web page scripts.


To answer \textbf{RQ1}, we align 12 trace pairs randomly taken from the initial set of all possible trace pairs ($600\times600$).
We compare STRAC with different implementations of DTW
1) From public github repositories: rmaestre \footnote{\url{ https://github.com/rmaestre/FastDTW}}, dtaidistance \footnote{\url{https://github.com/wannesm/dtaidistance}} and pierre-rouanet \footnote{\url{https://github.com/pierre-rouanet/dtw}}; 
2) From R's dtw package \cite{JSSv031i07} ;
3) The DTW implementation used in \cite{8621878}, slaypni \footnote{\url{https://github.com/slaypni/fastdtw}}.
For each comparison, we compute the average wall-clock execution time.

\textbf{RQ2} is answered as follows. We select a random sample of 100 pairs from all possible trace pairs ($600\times600$). We select 35 pairs of traces extracted from the same pages and 65 pairs of traces extracted from different pages.
Alignment cost is measured for each pair using gap cost $\gamma=1$ and event distance function $d_{Sen}$ (defined in \autoref{parametrization}), with parameters: $s=5$ and $c=0$. We group and plot each pair alignment cost per site.

We answer \textbf{RQ3} using the same traces as \textit{RQ2}. We compute DTW on each one of the 100 sampled pairs. We use the same gap cost $\gamma = 1$, but we compare the two distance functions $d_{Sen}$ and $d_{Inst}$ (defined in \autoref{parametrization}), with parameters: $s=5$ and $c=0$. We measure the alignment cost for each pair and compare the results with the ones obtained in \textit{RQ2}. 

The STRAC experimentation has been made on a PC with Intel Core i7 CPU and 16Gb DDR3 of RAM. We extract all traces from Chrome version 74.0.3729.169 (Official Build) (64-bit).

\label{pc_type}

\subsection{Answer to RQ1: Scalability}

\label{comparison}

\begin{figure}[htpb]
  \includegraphics[width=\linewidth]{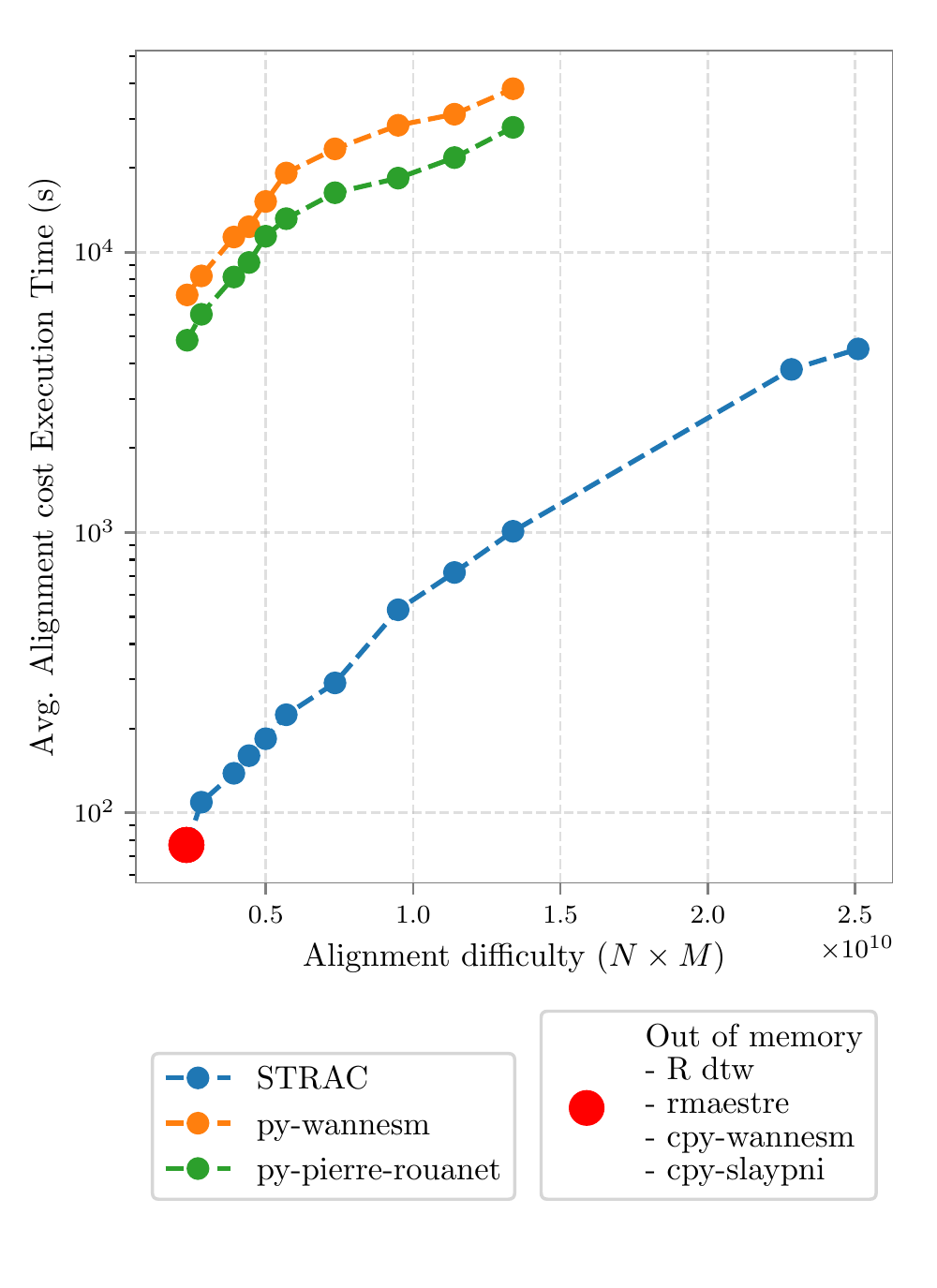}
  \caption{Execution time for \textbf{12} trace pair comparisons by 7 tools incl. STRAC. Y axis is in logarithmic scale. Four tools fail even on the smallest traces.}
  \label{times}
\end{figure}

\autoref{times} shows the execution time of 6 different alignment tools on 12 trace pairs.
The X axis gives the size of the alignment problem, which is the multiplication of the size of both traces in number of bytecode instructions.
The Y axis represents the execution time in seconds with a logarithmic scale.

First, we observe that four tools get out of memory for all the considered trace pairs: R-dtw, cpy-wannesm, rmaestre, cpy-slaypul (see the red dot in \autoref{times}). The main reason for this failure is that those tools need to store the cost matrix in memory. The least difficult trace comparison in the plot is a pair of traces of 48k instructions each. Finding the best alignment for this pair consists in analyzing an eight-bytes integer matrix of approx. 20GB (exactly 18632 millions of bytes). This memory requirement is almost the full memory of modern personal computers and it causes memory explosion at runtime. Applying the same analysis to the most difficult alignment in the plot shows requires 200GB of memory.


Second, py-wannesm and py-pierre-rouanet calculate the best alignment cost for the first 10 pairs, without any memory issue, even for problems in the order of magnitude close to $1.5\times 10^{10}$ in alignment difficulty. After this value, these tools also start to get memory issues for the same reason as the other tools. Yet, these succesfully align the 10 pairs (orange and green curves in \autoref{times}) thanks to an efficient use of Numpy \cite{numpy} arrays to store cost matrix. Numpy arrays in Python are tailored to efficiently deal with arrays up to 20GB of memory in x64 architectures. 
We also observe that  py-wannesm is always slower than py-pierre-rouanet. The main reason for this time difference is that py-wannesm does an extra pass through the cost matrix and py-pierre-rouanet does not do it.

Third, STRAC succesfully find the best alignment cost for all pairs in the benchmark, even for trace pairs that require memory beyond Numpy capabilities (the last two blue dots in \autoref{times}). The key insight behind is that STRAC implements the cost matrix data structure as a hybrid between memory and disk, i.e. moving such memory needs to disk.

Both Python implementations (py-wannesm and py-pierre-rouanet) systematically take at least one order of magnitude longer to run, compared to STRAC. The main reason behind this is that Python usually compiles code at runtime, while Java compiles it in advance, making a faster program. Besides, most JVMs perform Just-In-Time compilation to all or part of programs to native code, which significantly improves performance, but mainstream Python does not do this.

Recall that best alignment calculation using naive DTW implementation is non-scalable by its space-time quadratic nature, any implementation of DTW (even the one included in STRAC) eventually will run out of space (in memory or disk) and execution time will be near to impossible. However, STRAC can deal with all trace pairs of our benchmark thanks to its hybrid strategy that leverages both the disk and the memory. To align an average trace of 100k instructions, STRAC takes approx. 14 minutes in a PC like the one mentioned in \autoref{pc_type}.







\subsection{Answer to RQ2: Consistency}
\label{DTW_comparison}

\begin{figure*}
  \includegraphics[width=\textwidth]{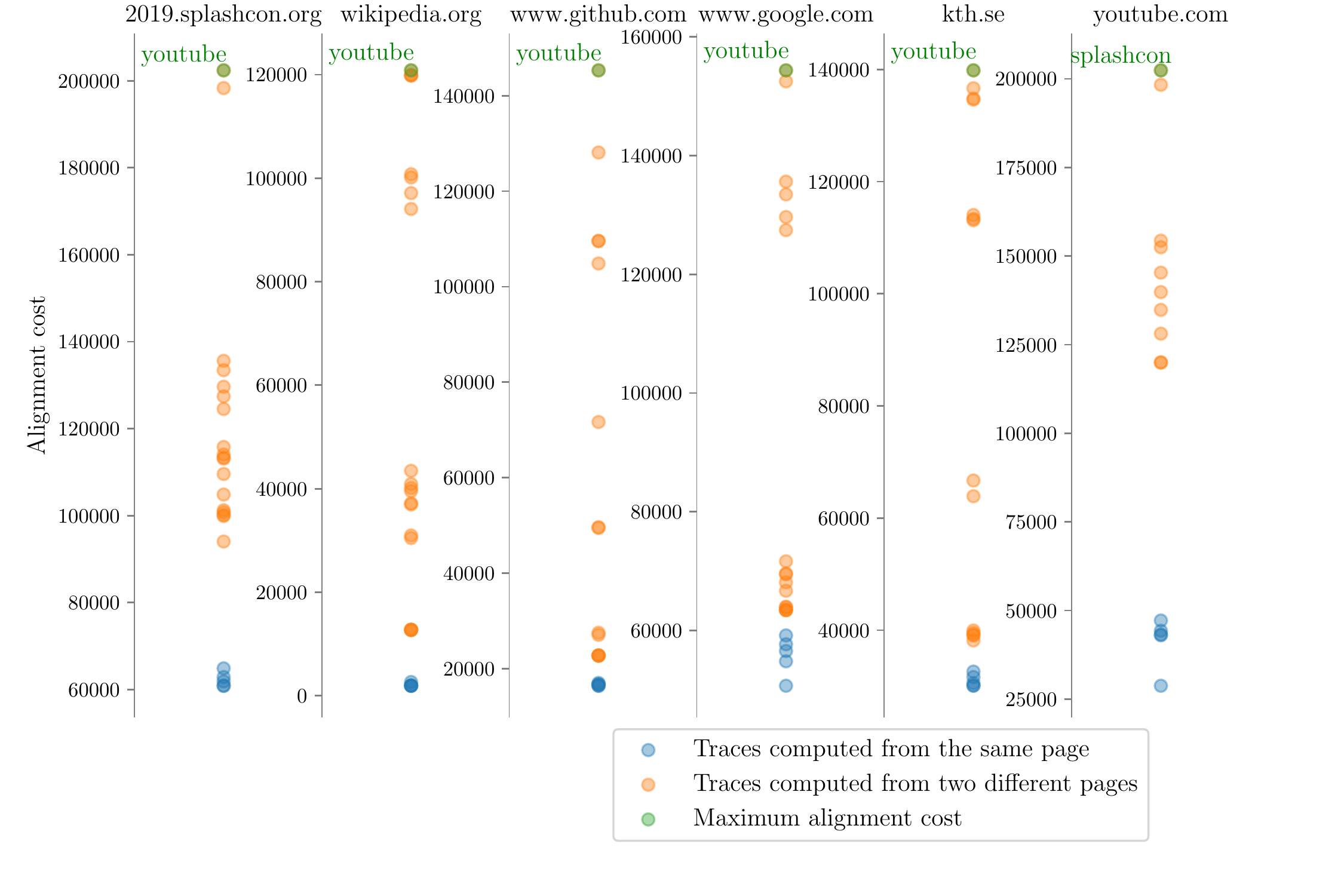}
  \caption{Alignment costs for 100 trace pair comparisons  using $d_{Sen}$ as distance function.}
  \label{convergence}
\end{figure*}

In \autoref{convergence}, we plot the alignment cost for 100 trace pairs, the blue dots represent pairs extracted from the same page, the orange dots illustrate trace pairs taken from two different pages. Each column corresponds to a given web page. Green dots represent pairs with the maximum alignment cost for each site: an alignment of  the web page treated in the column with a trace from the site cited above the dot. For example, the green dot in the first column is an alignment of a trace pair (2019.splashcon, youtube).

In  \autoref{convergence}, we observe that,  for each site, traces from the same page have a lower alignment cost. This is consistent with the fact that in these cases, the majority of both traces in the pair are the same.
On the contrary, the alignment cost between traces from different pages is higher.


Some cases show blue dots with sparsed high values. This occurs when  external scripts, declared in some pages, present a high variance in fetching process time. Also, it sometimes happens that for one script declared in a page, the remote servers sends different JavaScript code at each every request. Therefore, the generated bytecode varies more from one load to another, and the alignment cost is increased, showing a small margin between orange dots and the blue ones. However, we observe two scenarios when these phenomena are mitigated. First, when the bytecode generated from the external declaration is larger than the builtin bytecode (2019.splashcon, UNIV, and Youtube cases present a clear separation between clusters). Second, when the fetching process time is stable, as Wikipedia and Github cases show. 

In the case of Google, we observe the worst possible scenario. This site has 5 external declared scripts (see \autoref{sizes}),  3 of them have variable fetching time and their content varies at each load. These 3 scripts integrate Google Analytics features to the site. On the contrary, in the case of Wikipedia, external declared JavaScripts always provide the same code in almost constant time. As a result, the generated bytecode is more deterministic and  alignment cost decreases for traces from the same site. In the case of Wikipedia, alignment costs for pairs of traces collected from the same page vary between 1926 and 2652. These values are the lowest alignment costs in the benchmark, and they differ from others in more than $2\times$ in order of magnitude


Overall, the traces from the same (resp. different) page are located in separated clusters. In all cases, we also observe groups of orange dots that can be easily separated from other orange clusters. This separation is a consequence of semantic differences between sites and the increase of JavaScript declarations. For instance, in the first column of \autoref{convergence}, trace pairs from 2019.splashcon and Youtube home pages have higher alignment costs. This is a consequence of that Youtube is a richer feature site as 2019.splashcon is, but they semantically differ. We also observe this behavior in the case of Kth and Youtube trace pairs.




V8 compiles builtin JavaScript code to the same bytecode trace, as we discussed in \autoref{v8}. This bytecode generation is included in all collected traces. To validate this, we computed the V8 bytecode trace of an empty page: it contains 40k bytecode instructions on average. This also represents a constant noise in the alignment computation.

As  \autoref{convergence} illustrates, given the alignment cost of two semantically equivalent traces (blue dots) as a reference, STRAC is capable of identifying similarity with other page traces. However, we want to remark that STRAC accuracy gets improved when JavaScript declarations increase in the compared sites.



\subsection{Answer to RQ3: Distance Functions}
\label{similarity}

\begin{figure*}
  \includegraphics[width=\textwidth]{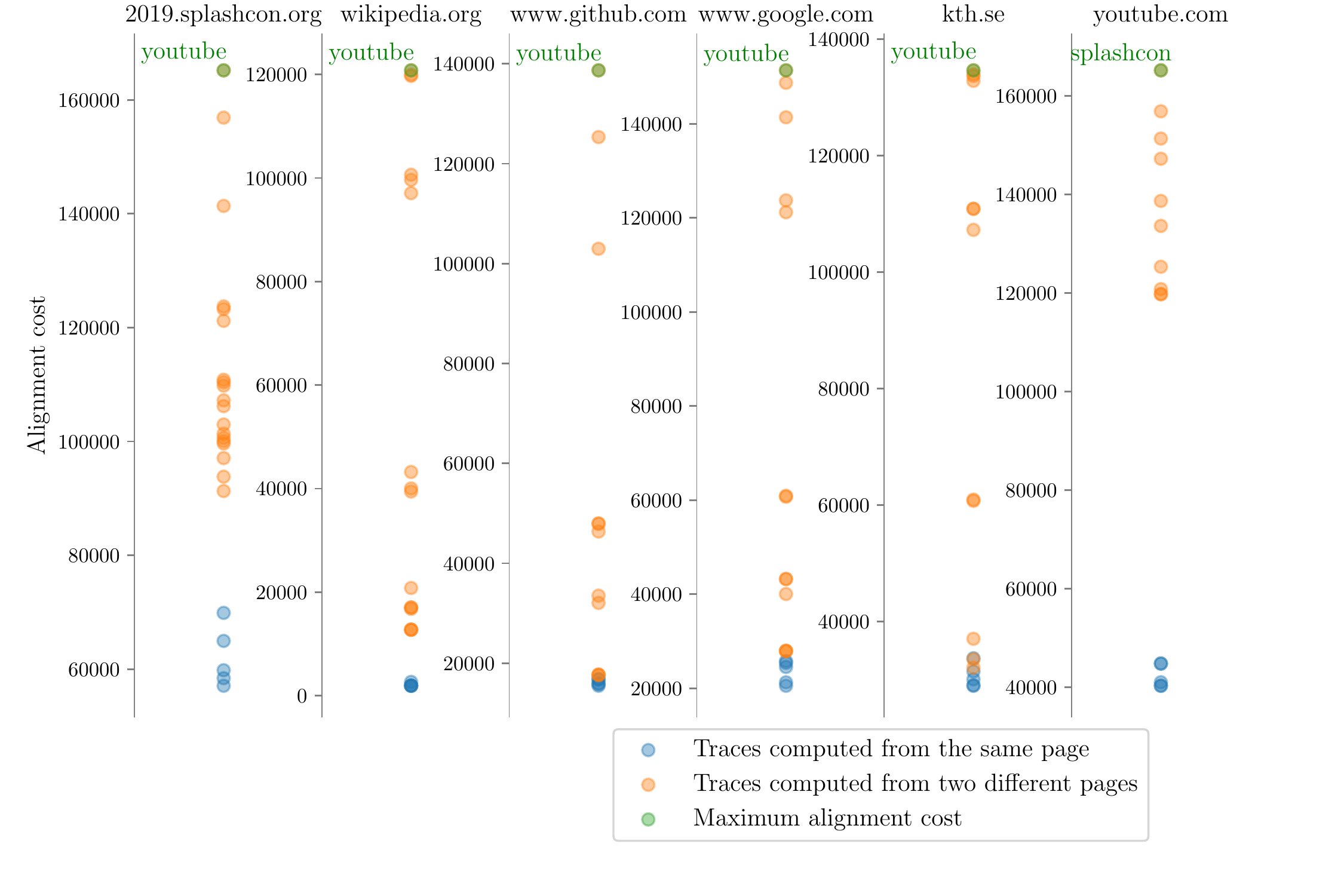}
  \caption{Alignment cost for 100 trace pair comparisons using $d_{Ins}$ as distance function.}
  \label{convergence2}
\end{figure*}

In \autoref{convergence2}, we plot the alignment cost using distance $d_{Ins}$. Recall that $d_{Ins}$ is less restrictive than $d_{Sen}$, the distance used to answer \textbf{RQ2}. 
By comparing \autoref{convergence2} and \autoref{convergence}, we observe interesting phenomena.
First, changing the distance function breaks the clustering breakdown for Github, Google and Kth (some blue points get mixed with orange points).
Second, the maximum alignment cost is lower than in \autoref{convergence} for all sites. These phenomena are consequences of using a less restrictive distance function, i.e. with $d_{Ins}$, only the operator is analyzed in the bytecode instructions comparison. 
Overall, the choice of distance function matters. 
STRAC can be extended with new distance functions and provides $d_{Sen}$ by default for properly aligning V8 bytecode traces.

We notice that the impact of the distance function is bigger for sites with less JavaScript.
For Google, Github and Wikipedia, using $d_{Ins}$ is bad because it breaks the clustering. For the remaining three websites, which involve more JavaScript features, while the alignment changes, the core property of the alignment of identifying semantically equivalent traces still holds.

\section{Related Work}
\label{rw}

DTW is memory greedy on trace size, a similar problem arises when dealing with streaming traces. Oregi et al. \cite{10.1007/978-3-319-71246-8_36} and Martins et al. \cite{8621878} present a generalization of DTW for large streaming data. They propose the use of incremental computation of the cost matrix complemented with a weighted event distance function adding event positions. However, their results may differ from the original DTW warp path. On the contrary, STRAC also computes the exact alignment cost without approximations. 

Kargen et al. \cite{Kargen:2017:TRI:3155562.3155608} propose a combination of data abstraction and FastDTW to align two program traces at the binary level. They record and analyze read and write operations to memory and x86 registers. Also, they argue and they show that their method scales to large traces. STRAC is also capable of analyzing such traces, but targets different kinds of traces: V8 bytecode traces, which are not handled by Kargen et al.

Ratanaworabhan et al. \cite{Ratanaworabhan:2010:JCB:1863166.1863169} instrument Internet Explorer to measure JavaScript runtime and static behavior in function calls and event handlers on real-world websites. By doing so, they show that common benchmarks, like SpiderMonkey and V8-Suite, are not representative of real application behavior. We could use STRAC to perform a similar analysis on modern browsers.  

With JALANGI, Sen et al. \cite{Sen:2013:JSR:2491411.2491447} provide a framework to dynamically analyze JavaScript. The framework works through source code instrumentation. JALANGI associates shadow values to variables and objects in the instrumented code, Sen et al. argue that most of  of state-of-the-art dynamic analysis techniques can be implemented, like concolic evaluation and taint analysis. However, JALANGI has several limitations dealing with builtin code and instrumentation can decrease instrumented code execution performance. With STRAC, we propose to use V8 bytecode traces to compare JavaScript semantic similarity without JavaScript instrumentation.

Fang et al. \cite{8482113} propose a JavaScript malicious code detection model based on neural networks. To mitigate the obfuscation techniques used in malicious code, they analyze the dynamic information recorded in V8 bytecode traces. Both STRAC and Fang et al. consider V8 bytecode traces, yet the usages are different: they do anomaly detection while we do trace comparison.

\section{Conclusion}
\label{conclusion}

In this paper, we presented a tool, called STRAC, for aligning execution traces. STRAC is tailored to traces of the JavaScript V8 engine. STRAC implements an optimized version of the DTW algorithm and two of its approximations. Our experiments show that STRAC scales to real-world JavaScript traces consisting of V8 bytecodes. STRAC provides two distance functions for trace event comparison and can be configured with any arbitrary distance function. Our evaluation indicates that STRAC performs better than state of the art DTW implementations, for 6 representative web sites.

We have shown that V8 bytecode contains redundancy and that an empty page includes more than 40k trace instructions. 
By removing this redundant and useless trace instructions, the alignment would get better. In our future work, we will study how to remove redundancy in V8 bytecode traces, for providing a better behavioral similarity measure for modern web pages full of JavaScript code.


\begin{acks}                            
This material is based upon work supported by the
{Swedish Foundation for Strategic Research} under the Trustfull project and by the  {Wallenberg Autonomous Systems and Software Program
(WASP)}.
\end{acks}

\balance
\bibliography{biblio.bib}


\begin{thebibliography}{29}


\ifx \showCODEN    \undefined \def \showCODEN     #1{\unskip}     \fi
\ifx \showDOI      \undefined \def \showDOI       #1{#1}\fi
\ifx \showISBNx    \undefined \def \showISBNx     #1{\unskip}     \fi
\ifx \showISBNxiii \undefined \def \showISBNxiii  #1{\unskip}     \fi
\ifx \showISSN     \undefined \def \showISSN      #1{\unskip}     \fi
\ifx \showLCCN     \undefined \def \showLCCN      #1{\unskip}     \fi
\ifx \shownote     \undefined \def \shownote      #1{#1}          \fi
\ifx \showarticletitle \undefined \def \showarticletitle #1{#1}   \fi
\ifx \showURL      \undefined \def \showURL       {\relax}        \fi
\providecommand\bibfield[2]{#2}
\providecommand\bibinfo[2]{#2}
\providecommand\natexlab[1]{#1}
\providecommand\showeprint[2][]{arXiv:#2}

\bibitem[\protect\citeauthoryear{Beschastnikh, Brun, Schneider, Sloan, and
  Ernst}{Beschastnikh et~al\mbox{.}}{2011}]%
        {beschastnikh_leveraging_2011}
\bibfield{author}{\bibinfo{person}{Ivan Beschastnikh}, \bibinfo{person}{Yuriy
  Brun}, \bibinfo{person}{Sigurd Schneider}, \bibinfo{person}{Michael Sloan},
  {and} \bibinfo{person}{Michael~D. Ernst}.} \bibinfo{year}{2011}\natexlab{}.
\newblock \showarticletitle{Leveraging Existing Instrumentation to
  Automatically Infer Invariant-Constrained Models}. In
  \bibinfo{booktitle}{\emph{Proceedings of the 19th {{ACM SIGSOFT}} Symposium
  and the 13th {{European}} Conference on {{Foundations}} of Software
  Engineering - {{SIGSOFT}}/{{FSE}} '11}} (2011). \bibinfo{publisher}{{ACM
  Press}}, \bibinfo{pages}{267}.
\newblock
\showISBNx{978-1-4503-0443-6}
\urldef\tempurl%
\url{https://doi.org/10.1145/2025113.2025151}
\showDOI{\tempurl}


\bibitem[\protect\citeauthoryear{Churchill, Padon, Sharma, and Aiken}{Churchill
  et~al\mbox{.}}{2019}]%
        {Churchill}
\bibfield{author}{\bibinfo{person}{Berkeley Churchill}, \bibinfo{person}{Oded
  Padon}, \bibinfo{person}{Rahul Sharma}, {and} \bibinfo{person}{Alex Aiken}.}
  \bibinfo{year}{2019}\natexlab{}.
\newblock \showarticletitle{Semantic Program Alignment for Equivalence
  Checking}. In \bibinfo{booktitle}{\emph{Proceedings of the 40th ACM SIGPLAN
  Conference on Programming Language Design and Implementation}}
  \emph{(\bibinfo{series}{PLDI 2019})}. \bibinfo{publisher}{ACM},
  \bibinfo{address}{New York, NY, USA}, \bibinfo{pages}{1027--1040}.
\newblock
\showISBNx{978-1-4503-6712-7}
\urldef\tempurl%
\url{https://doi.org/10.1145/3314221.3314596}
\showDOI{\tempurl}


\bibitem[\protect\citeauthoryear{community}{community}{2018}]%
        {numpy}
\bibfield{author}{\bibinfo{person}{Numpy community}.}
  \bibinfo{year}{2018}\natexlab{}.
\newblock \bibinfo{title}{Numeric python}.
\newblock
\newblock
\urldef\tempurl%
\url{https://www.numpy.org/index.html}
\showURL{%
\tempurl}


\bibitem[\protect\citeauthoryear{engine}{engine}{2016}]%
        {ignition}
\bibfield{author}{\bibinfo{person}{V8~JavaScript engine}.}
  \bibinfo{year}{2016}\natexlab{}.
\newblock \bibinfo{booktitle}{\emph{Ignition design documentation}}.
\newblock
\urldef\tempurl%
\url{https://v8.dev/docs/ignition}
\showURL{%
\tempurl}


\bibitem[\protect\citeauthoryear{{Fang}, {Huang}, {Liu}, and {Xue}}{{Fang}
  et~al\mbox{.}}{2018}]%
        {8482113}
\bibfield{author}{\bibinfo{person}{Y. {Fang}}, \bibinfo{person}{C. {Huang}},
  \bibinfo{person}{L. {Liu}}, {and} \bibinfo{person}{M. {Xue}}.}
  \bibinfo{year}{2018}\natexlab{}.
\newblock \showarticletitle{Research on Malicious JavaScript Detection
  Technology Based on LSTM}.
\newblock \bibinfo{journal}{\emph{IEEE Access}}  \bibinfo{volume}{6}
  (\bibinfo{year}{2018}), \bibinfo{pages}{59118--59125}.
\newblock
\showISSN{2169-3536}
\urldef\tempurl%
\url{https://doi.org/10.1109/ACCESS.2018.2874098}
\showDOI{\tempurl}


\bibitem[\protect\citeauthoryear{Giorgino}{Giorgino}{2009}]%
        {JSSv031i07}
\bibfield{author}{\bibinfo{person}{Toni Giorgino}.}
  \bibinfo{year}{2009}\natexlab{}.
\newblock \showarticletitle{Computing and Visualizing Dynamic Time Warping
  Alignments in R: The dtw Package}.
\newblock \bibinfo{journal}{\emph{Journal of Statistical Software, Articles}}
  \bibinfo{volume}{31}, \bibinfo{number}{7} (\bibinfo{year}{2009}),
  \bibinfo{pages}{1--24}.
\newblock
\showISSN{1548-7660}
\urldef\tempurl%
\url{https://doi.org/10.18637/jss.v031.i07}
\showDOI{\tempurl}


\bibitem[\protect\citeauthoryear{{Itakura}}{{Itakura}}{1975}]%
        {itakura_minimum_1975}
\bibfield{author}{\bibinfo{person}{F. {Itakura}}.}
  \bibinfo{year}{1975}\natexlab{}.
\newblock \showarticletitle{Minimum prediction residual principle applied to
  speech recognition}.
\newblock \bibinfo{journal}{\emph{IEEE Transactions on Acoustics, Speech, and
  Signal Processing}} \bibinfo{volume}{23}, \bibinfo{number}{1}
  (\bibinfo{date}{February} \bibinfo{year}{1975}), \bibinfo{pages}{67--72}.
\newblock
\showISSN{0096-3518}
\urldef\tempurl%
\url{https://doi.org/10.1109/TASSP.1975.1162641}
\showDOI{\tempurl}


\bibitem[\protect\citeauthoryear{{Jiang}, {Chen}, {Ungureanu}, and
  {Yoshihira}}{{Jiang} et~al\mbox{.}}{2007}]%
        {jiang_multiresolution_2007-1}
\bibfield{author}{\bibinfo{person}{G. {Jiang}}, \bibinfo{person}{H. {Chen}},
  \bibinfo{person}{C. {Ungureanu}}, {and} \bibinfo{person}{K. {Yoshihira}}.}
  \bibinfo{year}{2007}\natexlab{}.
\newblock \showarticletitle{Multiresolution Abnormal Trace Detection Using
  Varied-Length $n$-Grams and Automata}.
\newblock \bibinfo{journal}{\emph{IEEE Transactions on Systems, Man, and
  Cybernetics, Part C (Applications and Reviews)}} \bibinfo{volume}{37},
  \bibinfo{number}{1} (\bibinfo{date}{Jan} \bibinfo{year}{2007}),
  \bibinfo{pages}{86--97}.
\newblock
\showISSN{1094-6977}
\urldef\tempurl%
\url{https://doi.org/10.1109/TSMCC.2006.871569}
\showDOI{\tempurl}


\bibitem[\protect\citeauthoryear{{Kamiya}}{{Kamiya}}{2018}]%
        {8327321}
\bibfield{author}{\bibinfo{person}{T. {Kamiya}}.}
  \bibinfo{year}{2018}\natexlab{}.
\newblock \showarticletitle{Code difference visualization by a call tree}. In
  \bibinfo{booktitle}{\emph{2018 IEEE 12th International Workshop on Software
  Clones (IWSC)}}. \bibinfo{pages}{60--63}.
\newblock
\showISSN{2572-6587}
\urldef\tempurl%
\url{https://doi.org/10.1109/IWSC.2018.8327321}
\showDOI{\tempurl}


\bibitem[\protect\citeauthoryear{Karg{\'e}n and Shahmehri}{Karg{\'e}n and
  Shahmehri}{2017}]%
        {Kargen:2017:TRI:3155562.3155608}
\bibfield{author}{\bibinfo{person}{Ulf Karg{\'e}n} {and} \bibinfo{person}{Nahid
  Shahmehri}.} \bibinfo{year}{2017}\natexlab{}.
\newblock \showarticletitle{Towards Robust Instruction-level Trace Alignment of
  Binary Code}. In \bibinfo{booktitle}{\emph{Proceedings of the 32Nd IEEE/ACM
  International Conference on Automated Software Engineering}}
  \emph{(\bibinfo{series}{ASE 2017})}. \bibinfo{publisher}{IEEE Press},
  \bibinfo{address}{Piscataway, NJ, USA}, \bibinfo{pages}{342--352}.
\newblock
\showISBNx{978-1-5386-2684-9}
\urldef\tempurl%
\url{http://dl.acm.org/citation.cfm?id=3155562.3155608}
\showURL{%
\tempurl}


\bibitem[\protect\citeauthoryear{Kim, Kim, Kim, Kim, Kim, and Kim}{Kim
  et~al\mbox{.}}{2017}]%
        {Kim2017}
\bibfield{author}{\bibinfo{person}{Hyunjoo Kim}, \bibinfo{person}{Jonghyun
  Kim}, \bibinfo{person}{Youngsoo Kim}, \bibinfo{person}{Ikkyun Kim},
  \bibinfo{person}{Kuinam~J. Kim}, {and} \bibinfo{person}{Hyuncheol Kim}.}
  \bibinfo{year}{2017}\natexlab{}.
\newblock \showarticletitle{Improvement of malware detection and classification
  using API call sequence alignment and visualization}.
\newblock \bibinfo{journal}{\emph{Cluster Computing}} (\bibinfo{date}{12 Sep}
  \bibinfo{year}{2017}).
\newblock
\showISSN{1573-7543}
\urldef\tempurl%
\url{https://doi.org/10.1007/s10586-017-1110-2}
\showDOI{\tempurl}


\bibitem[\protect\citeauthoryear{Lemire}{Lemire}{2008}]%
        {DBLP:journals}
\bibfield{author}{\bibinfo{person}{Daniel Lemire}.}
  \bibinfo{year}{2008}\natexlab{}.
\newblock \showarticletitle{Faster Retrieval with a Two-Pass
  Dynamic-Time-Warping Lower Bound}.
\newblock \bibinfo{journal}{\emph{CoRR}}  \bibinfo{volume}{abs/0811.3301}
  (\bibinfo{year}{2008}).
\newblock
\showeprint[arxiv]{0811.3301}
\urldef\tempurl%
\url{http://arxiv.org/abs/0811.3301}
\showURL{%
\tempurl}


\bibitem[\protect\citeauthoryear{{Lou}, {Ao}, and {Dong}}{{Lou}
  et~al\mbox{.}}{2015}]%
        {7429637}
\bibfield{author}{\bibinfo{person}{Y. {Lou}}, \bibinfo{person}{H. {Ao}}, {and}
  \bibinfo{person}{Y. {Dong}}.} \bibinfo{year}{2015}\natexlab{}.
\newblock \showarticletitle{Improvement of Dynamic Time Warping (DTW)
  Algorithm}. In \bibinfo{booktitle}{\emph{2015 14th International Symposium on
  Distributed Computing and Applications for Business Engineering and Science
  (DCABES)}}. \bibinfo{pages}{384--387}.
\newblock
\urldef\tempurl%
\url{https://doi.org/10.1109/DCABES.2015.103}
\showDOI{\tempurl}


\bibitem[\protect\citeauthoryear{Maia, Sobreira, Paixão, Amo, and Silva}{Maia
  et~al\mbox{.}}{2008}]%
        {Maia08usinga}
\bibfield{author}{\bibinfo{person}{Marcelo De~A. Maia}, \bibinfo{person}{Victor
  Sobreira}, \bibinfo{person}{Klérisson~R. Paixão}, \bibinfo{person}{Ra~A.~De
  Amo}, {and} \bibinfo{person}{Ilmério~R. Silva}.}
  \bibinfo{year}{2008}\natexlab{}.
\newblock \showarticletitle{Using a sequence alignment algorithm to identify
  specific and common code from execution traces}. In
  \bibinfo{booktitle}{\emph{Proceedings of the 4th International Workshop on
  Program Comprehension through Dynamic Analysis (PCODA}}.
  \bibinfo{pages}{6--10}.
\newblock


\bibitem[\protect\citeauthoryear{{Martins} and {Kerren}}{{Martins} and
  {Kerren}}{2018}]%
        {8621878}
\bibfield{author}{\bibinfo{person}{R.~M. {Martins}} {and} \bibinfo{person}{A.
  {Kerren}}.} \bibinfo{year}{2018}\natexlab{}.
\newblock \showarticletitle{Efficient Dynamic Time Warping for Big Data
  Streams}. In \bibinfo{booktitle}{\emph{2018 IEEE International Conference on
  Big Data (Big Data)}}. \bibinfo{pages}{2924--2929}.
\newblock
\urldef\tempurl%
\url{https://doi.org/10.1109/BigData.2018.8621878}
\showDOI{\tempurl}


\bibitem[\protect\citeauthoryear{McIlroy}{McIlroy}{2016}]%
        {ignition16}
\bibfield{author}{\bibinfo{person}{Ross McIlroy}.}
  \bibinfo{year}{2016}\natexlab{}.
\newblock \bibinfo{title}{Ignition: V8 Interpreter}.
\newblock
\newblock
\urldef\tempurl%
\url{https://docs.google.com/document/d/11T2CRex9hXxoJwbYqVQ32yIPMh0uouUZLdyrtmMoL44/edit}
\showURL{%
\tempurl}


\bibitem[\protect\citeauthoryear{{Moreno}, {Treadway}, {Marcus}, and
  {Shen}}{{Moreno} et~al\mbox{.}}{2014}]%
        {6976081}
\bibfield{author}{\bibinfo{person}{L. {Moreno}}, \bibinfo{person}{J.~J.
  {Treadway}}, \bibinfo{person}{A. {Marcus}}, {and} \bibinfo{person}{W.
  {Shen}}.} \bibinfo{year}{2014}\natexlab{}.
\newblock \showarticletitle{On the Use of Stack Traces to Improve Text
  Retrieval-Based Bug Localization}. In \bibinfo{booktitle}{\emph{2014 IEEE
  International Conference on Software Maintenance and Evolution}}.
  \bibinfo{pages}{151--160}.
\newblock
\showISSN{1063-6773}
\urldef\tempurl%
\url{https://doi.org/10.1109/ICSME.2014.37}
\showDOI{\tempurl}


\bibitem[\protect\citeauthoryear{Needleman and Wunsch}{Needleman and
  Wunsch}{1970}]%
        {NEEDLEMAN1970443}
\bibfield{author}{\bibinfo{person}{Saul~B. Needleman} {and}
  \bibinfo{person}{Christian~D. Wunsch}.} \bibinfo{year}{1970}\natexlab{}.
\newblock \showarticletitle{A General Method Applicable to the Search for
  Similarities in the Amino Acid Sequence of Two Proteins}.
\newblock  \bibinfo{volume}{48}, \bibinfo{number}{3} (\bibinfo{year}{1970}),
  \bibinfo{pages}{443--453}.
\newblock
\showISSN{0022-2836}
\urldef\tempurl%
\url{https://doi.org/10.1016/0022-2836(70)90057-4}
\showDOI{\tempurl}


\bibitem[\protect\citeauthoryear{official~web page}{official~web page}{2019}]%
        {noauthor_v8_nodate-1}
\bibfield{author}{\bibinfo{person}{V8 official~web page}.}
  \bibinfo{year}{2019}\natexlab{}.
\newblock \bibinfo{booktitle}{\emph{V8 {{JavaScript}} Engine}}.
\newblock
\urldef\tempurl%
\url{https://v8.dev/}
\showURL{%
\tempurl}


\bibitem[\protect\citeauthoryear{Oregi, Pérez, Del~Ser, and Lozano}{Oregi
  et~al\mbox{.}}{2017}]%
        {10.1007/978-3-319-71246-8_36}
\bibfield{author}{\bibinfo{person}{Izaskun Oregi}, \bibinfo{person}{Aritz
  Pérez}, \bibinfo{person}{Javier Del~Ser}, {and} \bibinfo{person}{José~A.
  Lozano}.} \bibinfo{year}{2017}\natexlab{}.
\newblock \showarticletitle{On-Line Dynamic Time Warping for Streaming Time
  Series}. In \bibinfo{booktitle}{\emph{Machine Learning and Knowledge
  Discovery in Databases}}, \bibfield{editor}{\bibinfo{person}{Michelangelo
  Ceci}, \bibinfo{person}{Jaakko Hollmén}, \bibinfo{person}{Ljupco
  Todorovski}, {and} \bibinfo{person}{Saso Vens, Celinand~Dzeroski}} (Eds.).
  \bibinfo{publisher}{Springer International Publishing},
  \bibinfo{address}{Cham}, \bibinfo{pages}{591--605}.
\newblock
\showISBNx{978-3-319-71246-8}


\bibitem[\protect\citeauthoryear{Projects}{Projects}{2019}]%
        {noauthor_run_nodate}
\bibfield{author}{\bibinfo{person}{The~Chromium Projects}.}
  \bibinfo{year}{2019}\natexlab{}.
\newblock \bibinfo{booktitle}{\emph{Run {{Chromium}} with Flags - {{The
  Chromium Projects}}}}.
\newblock
\urldef\tempurl%
\url{https://www.chromium.org/developers/how-tos/run-chromium-with-flags#TOC-V8-Flags}
\showURL{%
\tempurl}


\bibitem[\protect\citeauthoryear{Ramos and Engler}{Ramos and Engler}{2011}]%
        {Ramos}
\bibfield{author}{\bibinfo{person}{David~A Ramos} {and}
  \bibinfo{person}{Dawson~R. Engler}.} \bibinfo{year}{2011}\natexlab{}.
\newblock \showarticletitle{Practical, Low-effort Equivalence Verification of
  Real Code}. In \bibinfo{booktitle}{\emph{Proceedings of the 23rd
  International Conference on Computer Aided Verification}}
  \emph{(\bibinfo{series}{CAV'11})}. \bibinfo{publisher}{Springer-Verlag},
  \bibinfo{address}{Berlin, Heidelberg}, \bibinfo{pages}{669--685}.
\newblock
\showISBNx{978-3-642-22109-5}
\urldef\tempurl%
\url{http://dl.acm.org/citation.cfm?id=2032305.2032360}
\showURL{%
\tempurl}


\bibitem[\protect\citeauthoryear{Ratanaworabhan, Livshits, and
  Zorn}{Ratanaworabhan et~al\mbox{.}}{2010}]%
        {Ratanaworabhan:2010:JCB:1863166.1863169}
\bibfield{author}{\bibinfo{person}{Paruj Ratanaworabhan},
  \bibinfo{person}{Benjamin Livshits}, {and} \bibinfo{person}{Benjamin~G.
  Zorn}.} \bibinfo{year}{2010}\natexlab{}.
\newblock \showarticletitle{JSMeter: Comparing the Behavior of JavaScript
  Benchmarks with Real Web Applications}. In
  \bibinfo{booktitle}{\emph{Proceedings of the 2010 USENIX Conference on Web
  Application Development}} \emph{(\bibinfo{series}{WebApps'10})}.
  \bibinfo{publisher}{USENIX Association}, \bibinfo{address}{Berkeley, CA,
  USA}, \bibinfo{pages}{3--3}.
\newblock
\urldef\tempurl%
\url{http://dl.acm.org/citation.cfm?id=1863166.1863169}
\showURL{%
\tempurl}


\bibitem[\protect\citeauthoryear{{Sakoe} and {Chiba}}{{Sakoe} and
  {Chiba}}{1978}]%
        {sakoe_dynamic_1978}
\bibfield{author}{\bibinfo{person}{H. {Sakoe}} {and} \bibinfo{person}{S.
  {Chiba}}.} \bibinfo{year}{1978}\natexlab{}.
\newblock \showarticletitle{Dynamic programming algorithm optimization for
  spoken word recognition}.
\newblock \bibinfo{journal}{\emph{IEEE Transactions on Acoustics, Speech, and
  Signal Processing}} \bibinfo{volume}{26}, \bibinfo{number}{1}
  (\bibinfo{date}{February} \bibinfo{year}{1978}), \bibinfo{pages}{43--49}.
\newblock
\showISSN{0096-3518}
\urldef\tempurl%
\url{https://doi.org/10.1109/TASSP.1978.1163055}
\showDOI{\tempurl}


\bibitem[\protect\citeauthoryear{Salvador and Chan}{Salvador and Chan}{2007}]%
        {Salvador:2007:TAD:1367985.1367993}
\bibfield{author}{\bibinfo{person}{Stan Salvador} {and} \bibinfo{person}{Philip
  Chan}.} \bibinfo{year}{2007}\natexlab{}.
\newblock \showarticletitle{FastDTW: Toward Accurate Dynamic Time Warping in
  Linear Time and Space}.
\newblock \bibinfo{journal}{\emph{Intell. Data Anal.}} \bibinfo{volume}{11},
  \bibinfo{number}{5} (\bibinfo{date}{Oct.} \bibinfo{year}{2007}),
  \bibinfo{pages}{561--580}.
\newblock
\showISSN{1088-467X}
\urldef\tempurl%
\url{http://dl.acm.org/citation.cfm?id=1367985.1367993}
\showURL{%
\tempurl}


\bibitem[\protect\citeauthoryear{Sen, Kalasapur, Brutch, and Gibbs}{Sen
  et~al\mbox{.}}{2013}]%
        {Sen:2013:JSR:2491411.2491447}
\bibfield{author}{\bibinfo{person}{Koushik Sen}, \bibinfo{person}{Swaroop
  Kalasapur}, \bibinfo{person}{Tasneem Brutch}, {and} \bibinfo{person}{Simon
  Gibbs}.} \bibinfo{year}{2013}\natexlab{}.
\newblock \showarticletitle{Jalangi: A Selective Record-replay and Dynamic
  Analysis Framework for JavaScript}. In \bibinfo{booktitle}{\emph{Proceedings
  of the 2013 9th Joint Meeting on Foundations of Software Engineering}}
  \emph{(\bibinfo{series}{ESEC/FSE 2013})}. \bibinfo{publisher}{ACM},
  \bibinfo{address}{New York, NY, USA}, \bibinfo{pages}{488--498}.
\newblock
\showISBNx{978-1-4503-2237-9}
\urldef\tempurl%
\url{https://doi.org/10.1145/2491411.2491447}
\showDOI{\tempurl}


\bibitem[\protect\citeauthoryear{Suzuki, Soares, Head, Glassman, Reis,
  Mongiovi, D'Antoni, and Hartmann}{Suzuki et~al\mbox{.}}{2017}]%
        {DBLP:journals/corr/abs-1708-03786}
\bibfield{author}{\bibinfo{person}{Ryo Suzuki}, \bibinfo{person}{Gustavo
  Soares}, \bibinfo{person}{Andrew Head}, \bibinfo{person}{Elena Glassman},
  \bibinfo{person}{Ruan Reis}, \bibinfo{person}{Melina Mongiovi},
  \bibinfo{person}{Loris D'Antoni}, {and} \bibinfo{person}{Bjoern Hartmann}.}
  \bibinfo{year}{2017}\natexlab{}.
\newblock \showarticletitle{TraceDiff: Debugging Unexpected Code Behavior Using
  Trace Divergences}.
\newblock \bibinfo{journal}{\emph{CoRR}}  \bibinfo{volume}{abs/1708.03786}
  (\bibinfo{year}{2017}).
\newblock
\showeprint[arxiv]{1708.03786}
\urldef\tempurl%
\url{http://arxiv.org/abs/1708.03786}
\showURL{%
\tempurl}


\bibitem[\protect\citeauthoryear{Verwaest and Hölttä}{Verwaest and
  Hölttä}{2019}]%
        {noauthor_blazingly_nodate}
\bibfield{author}{\bibinfo{person}{Toon Verwaest} {and} \bibinfo{person}{Marja
  Hölttä}.} \bibinfo{year}{2019}\natexlab{}.
\newblock \bibinfo{booktitle}{\emph{Blazingly Fast Parsing, Part 2: Lazy
  Parsing {$\cdot$} {{V8}}}}.
\newblock
\urldef\tempurl%
\url{https://v8.dev/blog/preparser}
\showURL{%
\tempurl}


\bibitem[\protect\citeauthoryear{{Weber}, {Brendel}, and {Brunst}}{{Weber}
  et~al\mbox{.}}{2012}]%
        {6280299}
\bibfield{author}{\bibinfo{person}{M. {Weber}}, \bibinfo{person}{R. {Brendel}},
  {and} \bibinfo{person}{H. {Brunst}}.} \bibinfo{year}{2012}\natexlab{}.
\newblock \showarticletitle{Trace File Comparison with a Hierarchical Sequence
  Alignment Algorithm}. In \bibinfo{booktitle}{\emph{2012 IEEE 10th
  International Symposium on Parallel and Distributed Processing with
  Applications}}. \bibinfo{pages}{247--254}.
\newblock
\showISSN{2158-9178}
\urldef\tempurl%
\url{https://doi.org/10.1109/ISPA.2012.40}
\showDOI{\tempurl}


\end{thebibliography}


\end{document}